\documentstyle[abe,overcite,np,12pt]{article}
\begin{document}
\input mssymb.tex
\pagestyle{empty}
\setlength{\baselineskip}{19.8pt}
\def\today{\ifcase\month\or
  January\or February\or March\or April\or May\or June\or
  July\or August\or September\or October\or November\or December\fi
  \ \ \number\year}
\vspace*{-50pt}
\rightline{\bf RIMS-1130}
\vspace*{40pt}
\centerline{\cmssB Anomaly Problem in a Simple Model Analogous to}
\vskip10pt
\centerline{\cmssB  the Lightcone-Gauge Two-Dimensional Quantum Gravity}
\vskip40pt
\centerline{
 Mitsuo Abe$^{\hbox{\sc a},\,}$\foot(1,{
  E-mail: abe@kurims.kyoto-u.ac.jp}) 
 and 
 Noboru Nakanishi$^{\hbox{\sc b},\,}$\foot(2,{
  E-mail: nbr-nakanishi@msn.com})}
\vskip15pt
\centerline{\it $^{\hbox{\sc a}}$Research Institute for Mathematical Sciences,
Kyoto University, Kyoto 606-01, Japan}
\vskip5pt
\centerline{\it $^{\hbox{\sc b}}$12-20 Asahigaoka-cho, Hirakata 573, Japan}
\vskip20pt
\centerline{ February\ \ 1997 }
\vskip80pt
\centerline{\bf Abstract}
\vskip5pt
It was found in the two-dimensional quantum gravity both in the de Donder 
gauge and in the lightcone gauge that one of field equations breaks down 
at the level of the representation, though the breakdown is very little.
It is shown that this anomalous behavior occurs also in a very simple model 
analogous to the lightcone-gauge two-dimensional quantum gravity.
This model, however, can be transformed into a free field theory by 
a nonsingular transformation.  Of course, the latter has no anomaly.  
The cause of the discrepancy is analyzed.
\vskip30pt
\noindent{{\it PACS:\/} 03.70.+k; 11.10.Kk}\hfill\break
\vfill\eject
\pagestyle{plain}
\Sec{Introduction}
Although the conventional covariant perturbation theory has been very 
useful in particle physics, it is {\it not adequate\/} to apply it to 
quantum gravity because to do so one must introduce a {\it wrong\/} 
assumption at the starting point \cite{N1}.
That is, it is {\it not adequate\/} to solve quantum gravity in the 
interaction picture.  We have therefore developed a new approach to 
the covariant quantum field theory for solving it {\it in the Heisenberg 
picture\/} [2--11]. 
The outline of our method is as follows.
\par
First, from field equations and equal-time commutation relations, 
we set up Cauchy problems for all full-dimensional commutators.  
We then solve them (if necessary, we expand them in powers of parameters 
involved).  
From this operator algebra, we calculate all multiple commutators.  
Finally, we construct all $n$-point Wightman functions so as to be 
consistent with all multiple commutators under the requirement of energy 
positivity and the generalized normal-product rule.  
The latter is necessary to define Wightman functions involving composite 
operators (i.e., products of field operators at the same spacetime point).
\par
Our method has been most successful in the two-dimensional quantum gravity 
in the de Donder gauge [3--7, 12--15]. 
We obtain a complete set of Wightman functions explicitly in the manifestly 
covariant and BRS-invariant way.  Our solution is quite satisfactory, 
but it exhibits anomaly in a field equation, though all right if it is 
once covariantly differentiated.
\par
Our method is applicable also to noncovariant theories.  
We have recently solved the two-dimensional quantum gravity 
in the lightcone gauge \cite{AN15},\foot(3,{This paper is referred to as 
I hereafter.})  which is a local version of Polyakov's 
``induced'' quantum gravity \cite{Pol}.  We have obtained the explicit 
solution, which satisfies Polyakov's Ward-Takahashi (WT) identities.  
Nevertheless, it exhibits anomaly: A field equation, which is a 
{\it second-order\/} differential equation, breaks down, while
Polyakov's equation, which follows from it and is a {\it third-order\/} 
differential equation, does not suffer from anomaly.
\par
In the present paper, we wish to analyze the problem of this new type of 
anomaly by using a very simple model.  This model is obtained by simplifying 
the lightcone-gauge two-dimensional quantum gravity discussed in I, 
and it again exhibits anomaly of the same type.  
Interestingly enough, however, 
{\it this model is equivalent to a free-field theory,\/}
that is, there is a nonsingular transformation of field operators connecting 
both theories.
Of course, a free-field theory is free of anomaly.  Thus we encounter a kind 
of dilemma.
We analyze its cause and find that the discrepancy is caused by the violation 
of the associative law for a product of distributions.
\par
The present paper is organized as follows.  
In Section 2, we briefly review the two-dimensional quantum gravity 
in the lightcone gauge.  
In Section 3, we propose a simple model analogous to it and show that 
this model 
has anomaly similar to the one found in the two-dimensional quantum gravity.
In Section 4, we point out that this model is equivalent to 
a free-field theory, which has no anomaly, and analyze why such a dilemma 
happens.  
In Section 5, we investigate the WT identities.
The final section is devoted to discussion. In the Appendix, we discuss 
the symmetry properties of our model.
\vskip50pt
\Sec{Review of the lightcone-gauge model}
We first consider the two-dimensional quantum gravity in the lightcone 
gauge.
Only one component of the gravitational field $g_{\mu\nu}$ survives; it is
denoted by $h$ ( $=h_{++}$ in Polyakov's notation ). Polyakov's Lagrangian
density is a nonlocal one \cite{Pol};  by introducing an auxiliary field
$\tb$,  it is rewritten into a local one.  With the lightcone coordinates
$x^+$ and $x^-$, the relevant Lagrangian density is written as
\begin{eqnarray}
&&\lag=\partial_-h\cdot\partial_-\tb-{1\over2}\alpha h(\partial_-\tb)^2
       +{1\over2}\gamma\alpha\partial_+\tb\cdot\partial_-\tb,
\end{eqnarray}
where $\gamma=1$ and $\alpha$ is an arbitrary nonzero real constant.
For convenience, we generalize $\gamma=1$ to $\gamma$ arbitrary.
\par
In I, we have completely solved the model defined by \eqno(2,1) 
by applying our method explained in the Introduction.  
In the following, we briefly review our main results.
\par
It is convenient to transform $\tb$ into $\rho$ by setting
\begin{eqnarray}
&& \tb={2\over\alpha}\log \rho.
\end{eqnarray}
Then field equations are as follows:
\begin{eqnarray}
&& \partial_-(\partial_-h-2\rho^{-1}\partial_-\rho\cdot h
              +2\gamma\rho^{-1}\partial_+\rho)=0, \\
&& \partial_-{}^2\rho=0.
\end{eqnarray}
It is easy to derive Polyakov's equation
\begin{eqnarray}
&& \partial_-{}^3h=0
\end{eqnarray}
from \eqno(2,3) with the help of \eqno(2,4).
It should be noted, however, that \eqno(2,3) {\it cannot\/} be reproduced
from \eqno(2,4) and \eqno(2,5).  Owing to \eqno(2,4) and 
\eqno(2,5), we may write
\begin{eqnarray}
&&\rho(x)=a(x^+)+c(x^+)x^-,\\
&&h(x)=j^{(1)}(x^+)-2j^{(0)}(x^+)x^-+j^{(-1)}(x^+)(x^-)^2.
\end{eqnarray}
\par
Regarding $x^-$ as the time variable, we can consistently carry out canonical
quantization.   Since the $x^-$ dependence is known as in \eqno(2,6) and 
\eqno(2,7), it is straightforward to calculate the two-dimensional 
commutators from the equal-time ones.  We find
\begin{eqnarray}
&&[\rho(x),\;\rho(y)]=0, \\
&&[h(x),\;\rho(y)]=-{1\over2}i\alpha(x^--y^-)\rho(x)\delta(x^+-y^+), \\
&&[h(x),\;h(y)]=-i\alpha(x^--y^-)h(x,y)\delta(x^+-y^+)
             +{1\over2}i\gamma\alpha(x^--y^-)^2\delta'(x^+-y^+), \nonumber\\
&&
\end{eqnarray}
where
\begin{eqnarray}
h(x,y)&=&h(x^-,\,y^-;\,x^+=y^+) \nonumber \\
      &=&j^{(1)}-j^{(0)}(x^-+y^-)+j^{(-1)}x^-y^- \nonumber \\
      &=&{1\over2}\Bigg[h(x)+h(y)
          -{x^--y^-\over2}\bigg(\partial_-h(x)-\partial_-h(y)\bigg)
                  \Bigg]_{x^+=y^+}.
\end{eqnarray}
\par
The operator solution \eqno(2,8)$\sim$\eqno(2,10) are uniquely 
characterized as the solutions to the Cauchy problems for two-dimensional
commutators, which are set up by using \eqno(2,3)$\sim$\eqno(2,5) 
together with the canonical commutation relations:
\begin{eqnarray}
&&\left\{\begin{array}{l}
         (\partial_-{}^x)^2[\rho(x),\;\rho(y)]=0, \\ \noalign{\vskip7pt}
          \phantom{(\partial_-{}^x)^2}[\rho(x),\;\rho(y)]|_{x^-=y^-}=0, 
            \\ \noalign{\vskip7pt}
          \phantom{()^2}\partial_-{}^x[\rho(x),\;\rho(y)]|_{x^-=y^-}=0; 
        \end{array}\right. \\ \noalign{\vskip7pt}
&&\left\{\begin{array}{l}
         (\partial_-{}^x)^2[\rho(x),\;h(y)]=0, \\ \noalign{\vskip7pt}
          \phantom{(\partial_-{}^x)^2}[\rho(x),\;h(y)]|_{x^-=y^-}=0, 
            \\ \noalign{\vskip7pt}
          \phantom{()^2}\partial_-{}^x[\rho(x),\;h(y)]|_{x^-=y^-}
            =-i\displaystyle{\alpha\over2}\rho\,\delta(x^+-y^+);
         \end{array}\right. \\ \noalign{\vskip7pt}
&&\left\{\begin{array}{l}
         (\partial_-{}^x)^3[h(x),\;h(y)]=0, \\ \noalign{\vskip7pt}
          \phantom{(\partial_-{}^x)^3}[h(x),\;h(y)]|_{x^-=y^-}=0,
            \\ \noalign{\vskip7pt}
          \phantom{()^3}\partial_-{}^x[h(x),\;h(y)]|_{x^-=y^-}
            =-i\alpha h \, \delta(x^+-y^+), \\ \noalign{\vskip7pt}
          (\partial_-{}^x)^2[h(x),\;h(y)]|_{x^-=y^-}
            =-i\alpha\partial_-h\cdot\delta(x^+-y^+)+i\gamma\delta'(x^+-y^+).
        \end{array}\right.
\end{eqnarray}
From \eqno(2,12)$\sim$\eqno(2,14), we can reproduce
\eqno(2,8)$\sim$\eqno(2,10) by using neither \eqno(2,6) 
nor \eqno(2,7). 
Indeed, \eqno(2,8) is the trivial solution to \eqno(2,12), while 
\eqno(2,9) and \eqno(2,10) are
shown to be solutions to \eqno(2,13) and \eqno(2,14), respectively, 
by using the following identities:
\begin{eqnarray}
 F(x,y)&=&{1\over4\pi}\int d^2u\,\ep(x,y;u)
          D(x-u)(\partial_-{}^u)^2F(u,y) \nonumber \\
       &&+{1\over2\pi}\int du^+\,\Bigg[
           D(x-y)\partial_-{}^uF(u,y) \nonumber \\
       && \hspace*{65pt}
          -\partial_-{}^uD(x-u)\cdot F(u,y)\Bigg]_{u^-=y^-}, \\
 G(x,y)&=&{1\over8\pi}\int d^2u\,\ep(x,y;u)
        \tilde D(x-u)(\partial_-{}^u)^3G(u,y) \nonumber \\
       &&+{1\over4\pi}\int du^+\,\Bigg[
           \tilde D(x-y)(\partial_-{}^u)^2G(u,y) \nonumber \\
       && \hspace*{65pt}
           -\partial_-{}^u\tilde D(x-u)\cdot\partial_-{}^u G(u,y) \nonumber \\
       && \hspace*{65pt}
           +(\partial_-{}^u)^2\tilde D(x-u)\cdot G(u,y)\Bigg]_{u^-=y^-}, 
\end{eqnarray}
where
\begin{eqnarray}
&& \ep(x,y;u)\equiv\theta(x^--u^-)-\theta(y^--u^-), \\
&& D(\xi)\equiv2\pi\xi^-\delta(\xi^+), \qquad \tilde D(\xi)\equiv\xi^-D(\xi).
\end{eqnarray}
\par
From \eqno(2,8)$\sim$\eqno(2,10), we calculate all multiple 
commutators explicitly.
\par
Assuming that 1-point functions $\wightman{\rho}$ and $\wightman{h}$ are
constants, we construct all Wightman functions so as to be consistent with 
all multiple commutators.
\par
Replacing $\Dp$ functions by $\Df$ functions, we obtain all $n$-point
$\tau$-functions (i.e., vacuum expectation values of time-ordered products).
Here the massless Feynman propagator $\Df$ is defined 
by\foot(4,{This definition is unconventional by a factor {$-(2\pi)^{-1}$}.})
\begin{eqnarray}
&&\Df(\xi)\equiv{\xi^-\over\xi^+-i0\xi^-}
=\xi^-\left({\theta(\xi^-)\over\xi^+-i0}+{\theta(-\xi^-)\over\xi^++i0}\right),
\end{eqnarray}
which satisfies
\begin{eqnarray}
&&(2\pi i)\partial_-{}^2\Df(\xi)=\delta^2(\xi).
\end{eqnarray}
\par
Polyakov's WT identities are derived by using \eqno(2,5) together with
the equal-time commutation relations and \eqno(2,20).  
It is explicitly confirmed that our solution satisfies them.
Indeed, our solution is completely consistent with \eqno(2,4) and 
\eqno(2,5).
It is {\it not\/}, however, consistent with \eqno(2,3).  
We thus encounter anomaly.
This anomaly cannot be eliminated by renormalizing the values of $\alpha$
and $\gamma$.
\vskip50pt
\Sec{A simple model}
The essential feature of the lightcone-gauge two-dimensional quantum gravity
is irrelevant to the presence of the last term of \eqno(2,1).  
Hence we may simplify the model by setting $\gamma=0$.  
Then \eqno(2,1) depends on $\tb$ only through 
$\partial_-\tb$.  Hence we can further simplify the model by replacing 
$\partial_-\tb$ by $b$, though this change is quite nontrivial.
\par
Thus we consider a simplified model given by the Lagrangian density
\begin{eqnarray}
&&\lag=b\partial_-h-{\alpha\over2}b^2h.
\end{eqnarray}
Field equations are
\begin{eqnarray}
&& \partial_-h-\alpha bh=0,\\
&& \partial_-b+{\alpha\over2}b^2=0.
\end{eqnarray}
\par
Setting $b=(2/\alpha)\rho^{-1}$, we reduce \eqno(3,3) to 
\begin{eqnarray}
&& \partial_-\rho=1,
\end{eqnarray}
whence we have
\begin{eqnarray}
&& \rho(x)=a(x^+)+x^-.
\end{eqnarray}
On the other hand, \eqno(3,2) is rewritten as
\begin{eqnarray}
&& \rho\partial_-h-2h=0.
\end{eqnarray}
Multiplying it by $\partial_-{}^2$ and using \eqno(3,4), we obtain 
\begin{eqnarray}
&& \partial_-{}^3h=0.
\end{eqnarray}
We can therefore express $h$ as in \eqno(2,7).  
In contrast to the lightcone-gauge model, however, $j^{(1)},\ j^{(0)}$ 
and $j^{(-1)}$ are not quite  independent in the present model.  
Indeed, substituting \eqno(3,5) and \eqno(2,7) into \eqno(3,6), 
we find
\begin{eqnarray}
&& j^{(0)}=-aj, \\
&& j^{(1)}=a^2j,
\end{eqnarray}
where $j\equiv j^{(-1)}$, so that
\begin{eqnarray}
&& h=\rho^2j.
\end{eqnarray}
\par
Canonical quantization\foot(5,{$b$ is not regarded as a canonical 
variable.})
can be carried out smoothly.  We then obtain
\begin{eqnarray}
&& [a(x^+),\;a(y^+)]=0, \\
&& [j(x^+),\;a(y^+)]=-i{\alpha\over2}\delta(x^+-y^+), \\
&& [j(x^+),\;j(y^+)]=0; \\
\noalign{\vskip5pt\noindent therefore, \vskip5pt}
&& [\rho(x),\;\rho(y)]=0, \\
&& [h(x),\;\rho(y)]=-i{\alpha\over2}\rho^2(x)\delta(x^+-y^+), \\
&& [h(x),\;h(y)]=-i\alpha[\rho(x)-\rho(y)]\rho(x)\rho(y)j\cdot\delta(x^+-y^+).
   \\
\noalign{\vskip5pt\noindent
Owing to \eqno(3,5) and \eqno(3,10), \eqno(3,16) is rewritten as 
\vskip5pt}
&& [h(x),\;h(y)]=-i\alpha(x^--y^-)h(x,y)\delta(x^+-y^+).
\end{eqnarray}
Thus $[h(x),\;h(y)]$ is formally the same as the $\gamma=0$ case of 
\eqno(2,10).
\par
Here, one should note that \eqno(3,14), \eqno(3,15) and 
\eqno(3,17) are directly obtained also
as the solutions to the Cauchy problems for two-dimensional commutators
by using neither \eqno(3,5) nor \eqno(3,10) in the same 
way as done in Section 2.
In this sense, the natural expression for $[h(x),\;h(y)]$ is 
\eqno(3,17), but not \eqno(3,16), although they are equivalent
at the level of operator solution.
\par
Nonvanishing multiple commutators are only those which involve at most one 
$\rho$.  Furthermore, as is seen from \eqno(3,17), the multiple 
commutators 
involving $h$'s only have the same expression as those in the $\gamma=0$ case
of the lightcone-gauge two-dimensional quantum gravity.  
The multiple commutators  involving one $\rho$ can be easily calculated by
using \eqno(3,15) and the above result.
\par
$1$-point functions are arbitrary.  For simplicity, we set
\begin{eqnarray}
&& \wightman{a}=0, \qquad \wightman{j}= \hbox{const}.
\end{eqnarray}
Then we have
\begin{eqnarray}
&&\wightman{\rho(x)}=x^-, \qquad \wightman{h(x)}=\wightman{j}(x^-)^2.
\end{eqnarray}
Thus {\it translational invariance is spontaneously broken\/}.   
This means that the conventional perturbative approach is 
{\it not applicable\/} to our model, because perturbation theory automatically
respects translational invariance as long as Feynman propagators are
translationally invariant.
\par
We present some of truncated Wightman functions (a subscript T indicates
truncation):
\begin{eqnarray}
&&\twightman{\rho(x_1)h(x_2)}
={\alpha\over4\pi}(x^-_2)^2\partial_-{}^{x_1}\Dp(x_1-x_2), \\
&&\twightman{h(x_1)h(x_2)}
=-{\alpha\over2\pi}\wightman{j}x^-_1x^-_2\Dp(x_1-x_2); \\
&&\twightman{\rho(x_1)h(x_2)h(x_3)}
={\alpha^2\over8\pi^2}
  \Bigg[x^-_2(x^-_3)^2\partial_-\Dp(x_1-x_2)\partial_-\Dp(x_2-x_3) \nonumber\\
&&\hspace*{150pt}
       -(x^-_2)^2x^-_3\partial_-\Dp(x_1-x_3)\partial_-\Dp(x_2-x_3)\Bigg],
  \hspace*{40pt} \\
&&\twightman{h(x_1)h(x_2)\cdots h(x_n)} \nonumber \\
&&\quad =\left(-{\alpha\over4\pi}\right)^{n-1}\wightman{j}
 \sum_{\hbox{\sc P}(i_1 \cdots i_n)}^{n!}x_{i_1}^-x_{i_n}^-
 \Dop(x_{i_1}-x_{i_2})\cdots  \Dop(x_{i_{n-1}}-x_{i_n}), 
\end{eqnarray}
where
\begin{eqnarray}
&&\Dp(\xi)={\xi^-\over\xi_+-i0} 
\end{eqnarray}
and $\Dop(x_i-x_j)$ equals $\Dp(x_i-x_j)$ for $i<j$ and $\Dp(x_j-x_i)$ for 
$i>j$ ; P$(i_1\, \cdots\, i_n)$ denotes a permutation of $(1,\, \ldots,\, n)$.
\par
Now, we proceed to the anomaly problem.
It is easy to confirm that all Wightman functions are consistent with 
\eqno(3,4) and \eqno(3,7).  
But the same is not true for \eqno(3,6).
\par
We calculate the {\it nontruncated\/}\foot(6,{For composite-operator 
calculation, nontruncated functions are necessary to be considered.}) 
function
\begin{eqnarray}
\wightman{\rho(x)\partial_-h(z)\cdot h(y)}
&=& \twightman{\rho(x)\partial_-h(z)\cdot h(y)}
 +\wightman{\rho(x)}\partial_-{}^z\twightman{h(z)h(y)}  \nonumber \\
&& +\partial_-\wightman{h(z)}\cdot\twightman{\rho(x)h(y)} 
   +\partial_-{}^z\twightman{\rho(x)h(z)}\wightman{h(y)} \nonumber \\
&& +\wightman{\rho(x)}\partial_-\wightman{h(z)}\cdot\wightman{h(y)}.
\end{eqnarray}
By using the generalized normal-product rule, which implies to drop all 
terms involving $\Dp(0)$ or its derivative, we find
\begin{eqnarray}
 \wightman{\rho(x)\partial_-h(x)\cdot h(y)}
&=&-{\alpha^2\over4\pi^2}x^-y^-[\partial_-\Dp(x-y)]^2 \nonumber \\
&& -2{\alpha\over2\pi}\wightman{j}x^-y^-\Dp(x-y)+2\wightman{j}^2(x^-)^2(y^-)^2,
\hspace*{30pt}
\end{eqnarray}
where use has been made of an identity
\begin{eqnarray}
&& \xi^-\partial_-\Dp(\xi)=\Dp(\xi).
\end{eqnarray}
The sum of the last two terms of \eqno(3,26) is equal to 
$2\wightman{h(x)h(y)}$.
Hence we have
\begin{eqnarray}
\wightman{[\rho(x)\partial_-h(x)-2h(x)]h(y)}
&=&-{\alpha^2\over4\pi^2}x^-y^-[\partial_-\Dp(x-y)]^2 \nonumber \\
&\not=& 0.
\end{eqnarray}
Thus the field equation \eqno(3,6) is violated at the level of 
representation.
\vskip50pt
\Sec{An equivalent free-field theory}
By partial integration, the Lagrangian density \eqno(3,1) is equivalent 
to 
\begin{eqnarray}
&&\tilde\lag=-(\partial_-b+{\alpha\over2}b^2)h.
\end{eqnarray}
We make the transformation
\begin{eqnarray}
&& b=(2/\alpha)\rho^{-1}, \qquad h=\rho^2j
\end{eqnarray}
at the Lagrangian level.  We then obtain
\begin{eqnarray}
&& \tilde\lag=(2/\alpha)(\partial_-\rho-1)j.
\end{eqnarray}
This is a free-field theory!   Furthermore, the Jacobian of the transformation
\eqno(4,2) is $-2/\alpha$; that is, it is {\it nonsingular\/}.  
Hence our model should be equivalent to a free-field theory.
\par
Field equations are
\begin{eqnarray}
&& \partial_-\rho=1, \qquad \partial_-j=0.
\end{eqnarray}
Two-dimensional commutators are the same as 
(3$\,\cdot\,$11)$\sim$(3$\,\cdot\,$13), that is,
\begin{eqnarray}
&& [j(x),\;\rho(y)]=-i{\alpha\over2}\delta(x^+-y^+)
\end{eqnarray}
and the others vanish.
\par
For a free-field theory, it is sufficient to give $1$-point and $2$-point
functions only.   In conformity with \eqno(3,18), $1$-point functions 
are set equal to 
\begin{eqnarray}
&& \wightman{\rho(x)}=x^-, \qquad \wightman{j(x)}=\wightman{j}.
\end{eqnarray}
The only nonvanishing truncated $2$-point function is 
\begin{eqnarray}
&& \twightman{\rho(x_1)j(x_2)}={\alpha\over4\pi}\partial_-\Dp(x_1-x_2).
\end{eqnarray}
\par
It is now straightforward to calculate Wightman functions involving $h(x)$, 
by replacing it as a composite field $\rho(x)^2j(x)$.  We then have
\begin{eqnarray}
\wightman{\rho(x)\partial_-h(x)\cdot h(y)}
&=&\wightman{\rho(x)\cdot 2\rho(x)\partial_-\rho(x)\cdot j(x)\rho^2(y)j(y)}
 \nonumber \\
&& +\wightman{\rho(x)\cdot\rho^2(x)\partial_-j(x)\cdot\rho^2(y)j(y)}.
\end{eqnarray}
Evidently, the first term of \eqno(4,8) equals $2\wightman{h(x)h(y)}$.
Since \eqno(4,6) and \eqno(4,7) are consistent with 
$\partial_-j=0$, the second term of \eqno(4,8) vanishes.  Thus we obtain
\begin{eqnarray}
&& \wightman{[\rho(x)\partial_-h(x)-2h(x)]h(y)}=0,
\end{eqnarray}
that is, we encounter {\it no anomaly\/}.
This result is, of course, inconsistent with \eqno(3,28). 
What is the cause of the discrepancy?
\par
We explicitly calculate $\wightman{h(x)h(y)}$ by regarding $h(x)$ as a
composite field $\rho^2(x)j(x)$ and by  applying the conventional 
normal-product rule.\foot(7,{When the generalized normal-product rule is
applied to a free-field theory, the conventional normal-product rule is
reproduced, as it should be.}) 
We then obtain
\begin{eqnarray}
\wightman{h(x)h(y)}&=&-{\alpha^2\over 4\pi^2}x^-y^-[\partial_-\Dp(x-y)]^2 
   \nonumber \\
                   && -{\alpha\over 2\pi}\wightman{j}x^-y^-\Dp(x-y)
                      +\wightman{j}^2(x^-)^2(y^-)^2.
\end{eqnarray}
Compared with the result obtained in Section 3, the first term of \eqno(4,10) 
is an {\it extra\/} contribution.  Thus the free-field theory is 
{\it different\/} from the model presented in Section 3 at the level of 
representation.
\par
From \eqno(4,10), we have
\begin{eqnarray}
\wightman{[h(x),\;h(y)]}&=& i{\alpha^2\over 2\pi}x^-y^-\delta'(x^+-y^+) 
 \nonumber \\
                        &&-i\alpha\wightman{j}x^-y^-(x^--y^-)\delta(x^+-y^+).
\end{eqnarray}
Is this formula consistent with \eqno(3,16)?
Evidently, if $x^+=y^+$ is executed for the operator product, 
we obtain the last term of \eqno(4,11) only, as was done in Section 3.  
But if we postpone the execution of $x^+=y^+$ for a moment, the 
calculation in terms of $\rho$ and $j$ yields
\begin{eqnarray}
\wightman{[h(x),\;h(y)]}&=&[\wightman{\rho(x)j(x)\rho^2(y)}
 - (\ x\ \leftrightarrow\ y \ )]\cdot i\alpha\delta(x^+-y^+) \nonumber \\
&=& -i{\alpha^2\over2\pi}x^-y^-
  \Bigg[{1\over x^+-y^+-i0}-{1\over y^+-x^+-i0}\Bigg]\delta(x^+-y^+)\nonumber\\
&&  -i\alpha\wightman{j}x^-y^-(x^--y^-)\delta(x^+-y^+).
\end{eqnarray}
Since the quantity in the square bracket equals 
$\displaystyle{2{\hbox{P}\over x^+-y^+}}$, where P denotes Cauchy principal
value, \eqno(4,12) reproduces \eqno(4,11) if we use the formula
\begin{eqnarray}
{\hbox{P}\over \xi}\cdot \delta(\xi)=-{1\over 2}\delta'(\xi).
\end{eqnarray}
Thus the presence or absence of the first term of \eqno(4,11) is 
the consequence of the violation of the associative law in the product 
of distributions.
\par
The above result is reconfirmed by considering the $c$-number term 
in the commutator $[\,:\!\!\rho^2(x)j(x)\!\!:,\; :\!\!\rho^2(y)j(y)\!\!:\,]$, 
where $:\quad:$ denotes the conventional normal product for free fields. 
One should note, however, that \eqno(4,11) is no longer consistent with 
the equal-time commutation relations for $[h(x),\,h(y)]|_{x^-=y^-}$ and 
$\partial_-{}^x[h(x),\,h(y)]|_{x^-=y^-}$ [see \eqno(2,14)] 
in the system of $h$ and $\rho$ described in Section 3.
A well-known example of this type of pathology is the calculation of the
Schwinger term based on the canonical anticommutation relations.
\par
The above justification of the presence of the first term of \eqno(4,11) 
is, of course, heavily based on the specialty of the free-field theory.
It is impossible to extend such a prescription to more general framework.
\vfill\eject
\Sec{Ward-Takahashi identities}
In this section, we discuss the $\tau$-functions.
They satisfy the WT identities, which are derived from \eqno(3,4) and 
\eqno(3,7) together with equal-time commutators.  
Therefore, their validity is independent of the anomaly problem.  
On the other hand, we have found two
different sets of $n$-point functions in Section 3 and in Section 4.  
Is it possible that both sets satisfy the same WT identities?
\par
It is straightforward to derive the following WT identities:
\begin{eqnarray}
&&\hspace*{-20pt}\ctau{\rho(x_1)h(x_2)\cdots h(x_n)} \nonumber\\
&&\hspace*{-20pt}
\quad ={\alpha\over4\pi}\sum_{k=2}^n\partial_-{}^{x_1}\Df(x_1-x_k)
   \ctau{\rho^2(x_k)h(x_2)\cdots\widehat{h(x_k)}\cdots h(x_n)},\\
\noalign{\vskip5pt\noindent with \vskip5pt}
&&\hspace*{-20pt}\ctau{\rho^2(x_1)h(x_2)\cdots h(x_n)} \nonumber\\
&&\hspace*{-20pt}\quad =\sum_{\hbox{\sc div}}\ctau{\rho(x_1)h(x_{i_1})\cdots}
                        \ctau{\rho(x_1)h(x_{\ell_1})\cdots}, \\
&&\hspace*{-20pt}\ctau{h(x_1)h(x_2)\cdots h(x_n)} \nonumber\\
&&\hspace*{-20pt}\quad =
-{\alpha\over4\pi}\sum_{k=2}^n\Df(x_1-x_k)[(x^-_1-x^-_k)\partial_-{}^{x_k}+2]
  \ctau{h(x_2)\cdots h(x_n)} \quad \hbox{for }n\geqq2, 
\hspace*{40pt}
\end{eqnarray}
where T, C and $\widehat{\ \ }$ denote time-ordered product, connected part 
and omission, respectively, and $\{\,(i_1,\,\ldots),\,(\ell_1,\,\ldots)\,\}$ 
is a partition of $(2,\,\ldots,\,n)$.  The derivation of \eqno(5,3) is 
the same as in I.
\par
As for $\ctau{\rho h \cdots h}$, there is no problem; both sets of the 
$\tau$-functions are the same.  Hence we concentrate our attention to 
$\ctau{h h \cdots h}$.
\par
From \eqno(3,23), we have 
\begin{eqnarray}
&&\ctau{h(x_1)h(x_2)\cdots h(x_n)}{}^{(h,\,\rho)}\nonumber \\
&&=\left(-{\alpha\over4\pi}\right)^{n-1}\wightman{j}
 \sum_{\hbox{\sc P}(i_1\,\cdots\,i_n)}^{n!} 
 x^-_{i_1}x^-_{i_n}\Df(x_{i_1}-x_{i_2})\cdots \Df(x_{i_{n-1}}-x_{i_n}),
\end{eqnarray}
where a superscript $(h,\,\rho)$ indicates that the calculation is made on the
basis of \eqno(3,1).
\par
We can prove that \eqno(5,4) satisfies \eqno(5,3).  
The proof is done essentially in 
the same way as done in I.  The only new situation is the presence of the 
end-point factors $x^-_{i_1}x^-_{i_n}$.  If $x_j$ is an end point, 
we encounter a factor
\begin{eqnarray}
&& [(x^-_1-x^-_j)\partial_-{}^{x_j}+1]x^-_j=x^-_1
\end{eqnarray}
[The square-bracket factor of \eqno(5,5) corresponds to the second part 
of (7$\,\cdot\,$12) of I in the end-point case.]  
That is, the old end-point factor $x^-_j$ is 
replaced by the new one $x^-_1$, as it should be.
\par
Now, we consider the $\tau$-functions resulting from the free-field theory 
\eqno(4,3), to which we attach a superscript $(j,\,\rho)$.  
We can easily see that
\begin{eqnarray}
&&\ctau{h(x_1)\cdots h(x_n)}{}^{(j,\,\rho)}
=\ctau{h(x_1)\cdots h(x_n)}{}^{\hbox{\sc tree}}
+\ctau{h(x_1)\cdots h(x_n)}{}^{\hbox{\sc loop}}  \nonumber \\
&& \hspace*{300pt} \hbox{ for } n\geqq2, 
\end{eqnarray}
with
\begin{eqnarray}
&&\ctau{h(x_1)\cdots h(x_n)}{}^{\hbox{\sc tree}}
 =\sum_{k=1}^n\wightman{j}\ctau{\rho^2(x_k)h(x_1)\cdots\widehat{h(x_k)}\cdots 
                                 h(x_n)}, \\
&&\ctau{h(x_1)\cdots h(x_n)}{}^{\hbox{\sc loop}} \nonumber \\
&&\qquad =-\left({\alpha\over 2\pi}\right)^n\prod_{k=1}^n x^-_k
   \sum_{\hbox{\sc C}(i_1\,\cdots\,i_n)}^{(n-1)!}
    \partial_-\Df(x_{i_1}-x_{i_2})\cdots\partial_-\Df(x_{i_n}-x_{i_1}),
\end{eqnarray}
where $C(i_1\,\cdots\,i_n)$ denotes a permutation of cyclic ordering 
$(1,\,2,\,\ldots,\,n)$.\foot(8,{In I, ``cyclic permutation'' should be 
read as ``permutation of cyclic ordering''.})
Diagrammatically, \eqno(5,7) is a sum over the contributions of the 
tree graphs in which no vertex has the degree more than 3.
Because of \eqno(4,7), the propagator is $\partial_-\Df$ but not $\Df$ 
in contrast to \eqno(5,4).  
By explicit calculation, we have confirmed that
\begin{eqnarray}
&&\ctau{h(x_1)\cdots h(x_n)}{}^{\hbox{\sc tree}}
  =\ctau{h(x_1)\cdots h(x_n)}{}^{(h,\,\rho)}
\end{eqnarray}
for $n=1,\,2,\,3.$  It is quite likely that \eqno(5,9) holds for $n$ 
general.
\par
The WT identity \eqno(5,3) shows that the number of propagators 
increases by one as $n$ increases by one.   
Accordingly, if $\ctau{h\cdots h}{}^{(j,\,\rho)}$
satisfies \eqno(5,3), then both $\ctau{h\cdots h}{}^{\hbox{\sc tree}}$ 
and $\ctau{h\cdots h}{}^{\hbox{\sc loop}}$ must satisfy it separately.
Since the former is all right provided that \eqno(5,9) is true, 
we have only to consider the latter.
\par
For $n=2$, \eqno(5,8) becomes
\begin{eqnarray}
\ctau{h(x_1)h(x_2)}{}^{\hbox{\sc loop}}=-\left({\alpha\over2\pi}\right)^2
x^-_1x^-_2[\partial_-\Df(x_1-x_2)]^2.
\end{eqnarray}
Since we must set $\ctau{h(x_1)}{}^{\hbox{\sc loop}}=0$, \eqno(5,10) 
should vanish if it obeys the WT identity \eqno(5,3).  
This is evidently impossible.
Higher-point functions also do not satisfy \eqno(5,3).  
\par
We thus conclude that {\it the free-field theory violates the WT identities\/}.
This result is understandable as a consequence of the violation of the 
equal-time commutation relations for $h$ in the free-field theory.
\vskip50pt
\Sec{Discussion}
In the present paper, we have investigated a simple model analogous to the 
lightcone-gauge two-dimensional quantum gravity.  In spite of its simplicity,
the former exhibits the anomaly of the same type as in the latter.
It is very interesting, however, that our model can be transformed, by a 
nonsingular transformation, into a free-field theory, which is free of anomaly.
We have analyzed the cause of the discrepancy and found that {\it it is the 
violation of the associative law for the product of distributions\/} as is 
encountered also in the analysis of the Schwinger term.
\par
We have considered the WT identities of our model, and found that the 
free-field theory does not satisfy them.   Accordingly, we should conclude
that {\it the free-field theory cannot be regarded as the anomaly-free 
version of the original model.\/}
\par
Two-dimensional massless field theories are, in general, very peculiar models.
It is not clear whether or not the pathological phenomena which we have found
are characteristic to two-dimensional massless field theories.  
It is desirable to make further investigation by means of a variety of models.
\vskip50pt
\setcounter{eqn}{0}
\renewcommand{\theequation}{\addtocounter{eqn}{1} 
               \rm $\!$A.\arabic{eqn}$\,$}
\leftline{\bf Appendix. \  Symmetry properties}
\vskip5pt
Polyakov \cite{Pol} found an $SL(2,\hbox{\bfit R})$ current algebra based on
$j^{(1)}(x^+)$, $j^{(0)}(x^+)$ and $j^{(-1)}(x^+)$ in his ``induced'' quantum
gravity.  This result can be understood as the $x^+$-dependent 
$SL(2,\hbox{\bfit R})$ symmetry whose generators are essentially given by 
$j^{(1)}$, $j^{(0)}$ and $j^{(-1)}$.  In our local version of the model,
we have found the existence of an extremely huge $x^+$-dependent 
symmetry \cite{AN16}; Polyakov's $SL(2,\hbox{\bfit R})$ is no more than a very
very tiny subalgebra of it.  Since our model considered in Section 3 is 
analogous to the lightcone-gauge two-dimensional quantum gravity, we can 
find the existence of a similar huge $x^+$-dependent symmetry.
\par
Because of \eqno(3,8) and \eqno(3,9), $j^{(0)}$ and $j^{(1)}$ are 
no longer independent
of $a$ and $j\equiv j^{(-1)}$.  The generator $Q$ of our huge algebra is 
formed from $a$ and $j$ only:
\def\bolda{\hbox{\bfit a}} \def\boldj{\hbox{\bfit j}}
\begin{eqnarray}
&&Q={2\over\alpha}\int dx^+\,F(x^+,\,\bolda,\,\boldj),
\end{eqnarray}
where $F$ is an arbitrary function, and $\bolda$ denotes a set of 
$a\equiv \rho-x^-$ and its finite-order $x^+$-derivatives, $\boldj$ being
similar.   By the generator $Q$, $a$ and $j$ are transformed as follows:
\begin{eqnarray}
&&\delta a\equiv i\vep[Q,\;a]=+\vep\left({\delta\over\delta\boldj}\right)_+F,\\
&&\delta j\equiv i\vep[Q,\;j]=-\vep\left({\delta\over\delta\bolda}\right)_+F,\\
\noalign{\vskip5pt \noindent where \vskip5pt}
&& \left({\delta\over\delta\bolda}\right)_+F\equiv\sum_{\ell}(-1)^\ell
   \partial_+{}^\ell\left({\partial F\over \partial(\partial_+^\ell a)}\right).
\end{eqnarray}
Under this transformation, the action $\int d^2x\,\lag=\int d^2x\,\tilde\lag$
is invariant because
\begin{eqnarray}
\delta\tilde\lag&=&
 \vep{2\over\alpha}\partial_-\Bigg[-F
  +j\left({\delta\over\delta\boldj}\right)_+F \Bigg] \nonumber\\
&&+\vep{2\over\alpha}\partial_+\Bigg[\sum_{\vphi=a,\,j}\sum_\ell\sum_{m=1}^\ell
   (-1)^{m+1}\partial_+{}^{\ell-m}\partial_-\vphi\cdot
   \partial_+{}^{m-1}{\partial F\over \partial(\partial_+^\ell\vphi)}\Bigg].
\end{eqnarray}
The Noether current $J^\mu$ is given by
\begin{eqnarray}
&&J^-={2\over\alpha}F, \qquad J^+=0,
\end{eqnarray}
in which use has been made of $\partial_-a=\partial_-j=0$.
\vfill\eject
\end{document}